\documentclass[conference]{IEEEtran}
\IEEEoverridecommandlockouts
\usepackage{cite}
\usepackage{amsmath,amssymb,amsfonts}
\usepackage[utf8]{inputenc}
\usepackage{afterpage}
\usepackage{graphicx,float}
\usepackage{placeins}
\usepackage{caption}
\usepackage{subcaption}
\usepackage{textcomp}
\usepackage{cite}
\usepackage{url}
\usepackage{listings}
\usepackage{xcolor}
\def\BibTeX{{\rm B\kern-.05em{\sc i\kern-.025em b}\kern-.08em
    T\kern-.1667em\lower.7ex\hbox{E}\kern-.125emX}}
\begin{document}
\title{BUSU-Net: An Ensemble U-Net Framework for Medical Image Segmentation}

\author{\IEEEauthorblockN{Wei Hao Khoong}
\IEEEauthorblockA{\textit{Department of Statistics and Applied Probability} \\
\textit{National University of Singapore}\\
khoongweihao@u.nus.edu}
}

\maketitle

\begin{abstract}
In recent years, convolutional neural networks (CNNs) have revolutionized medical image analysis. One of the most well-known CNN architectures in semantic segmentation is the U-net, which has achieved much success in several medical image segmentation applications. Also more recently, with the rise of autoML ad advancements in neural architecture search (NAS), methods like NAS-Unet have been proposed for NAS in medical image segmentation. In this paper, with inspiration from LadderNet, U-Net, autoML and NAS, we propose an ensemble deep neural network with an underlying U-Net framework consisting of bi-directional convolutional LSTMs and dense connections, where the first (from left) U-Net-like network is deeper than the second (from left). We show that this ensemble network outperforms recent state-of-the-art networks in several evaluation metrics, and also evaluate a lightweight version of this ensemble network, which also outperforms recent state-of-the-art networks in some evaluation metrics. 
\end{abstract}


\section{Introduction}

In the field of medical imaging, medical image segmentation has been playing an increasingly important role over the years. There has been an increasing demand for accurate, fast and cost-effective automated processing in medical image analysis equipment, such as Computed Tomography (CT), Magnetic Resonance Imaging (MRI) and ultrasound. Automated processes can not only save time and costs, but also reduce the reliance on manual labor (radiographer) and human error. 

To provide grounds for clinical diagnosis and to assist the doctors in making more accurate diagnoses, the medical imaging analysis equipment has to segment portions of interest (e.g. diseased vessels) in the medical image and extract relevant features out of it. These can be done via deep learning methods such as convolutional neural networks (CNNs), which is a booming area of active research in recent years. Research and applications on deep learning methods for image segmentation, alongside big data and cloud computing has been driving significant progress in the field of computer vision. Deep neural networks are mostly applied to classification problems, where the output of the network is in the form of a single label or a set of probability values linked to a given input image. Fully convolutional neural networks (FCN) was one of the first deep network method applied to image segmentation. The FCN architecture was extended to U-Net~\cite{olaf2015}, which achieved state-of-the-art results in segmentation without the need for a large amount of data. This not only saves time during training of the model, but also opens up new directions of research which can leverage the minimal requirement of image data to produce more accurate and robust image segmentation networks. 

\section{Related Work}

In medical imaging, semantic segmentation has playing an important role in recent years. Deep learning approaches have been achieving performance almost equivalent (or better) than that of radiologists, which in many use cases, have helped reduce the manual labor required for segmentation and also improved the speed and accuracy of the segmentations. Recent advancements in deep learning approaches for semantic segmentation include the fully convolutional neural network~\cite{john2015} (CNN), U-Net~\cite{olaf2015} which obtained the highest accuracy in the segmentation of neuronal structures in electron microscopic stacks, V-Net~\cite{milletari2016} which is a 3D extension of U-Net to predict segmentation of a input volume all at once, VoxResNet~\cite{chen2018} which is a deep voxel-wise residual network proposed for brain segmentation in MR, and NAS-Unet~\cite{nasunet2019}, a neural architecture search framework for semantic segmentation inspired by the U-net architecture. 

\subsection{LadderNet}

LadderNet~\cite{laddernet}, a multi-branch CNN with a chain of U-nets for semantic segmentation, contains skip connections between levels of the network and unlike the U-Net, its features from encoder branches are concatenated with features from decoder branches, in which features are summed from two branches. In particular, the authors propose LadderNet as an ensemble of multiple fully-convolutional networks, similar to Veit et al.~\cite{veit2016} who proposed that the ResNet performs like an ensemble of shallow networks, since the residual connections provided multiple paths of information flow. Here, every path can be viewed as a variant of a fully-convolutional network, and the total number of paths grow exponentially with the number of encoder-decoder pairs and spatial levels. This indicates that LadderNet has the potential to record more complicated features and yield higher accuracies. 

As the number of encoder-decoder pairs grow, so will the number of parameters and difficulty of training. The authors proposed a shared-weights residual block between ensembles which reaps the benefits of skip connections, recurrent convolutions, dropout regularization and also has much fewer parameters than a standard residual block. 

\subsection{BCDU-Net}

BCDU-Net~\cite{azad2019} was proposed as an extension of U-Net, and it yielded better performance than state-of-the-art alternatives for the task of segmentation. In its architecture, the contracting path consists of four steps, where each step has two convolutional $3\times 3$ filters followed by a $2\times 2$ max-pooling and a ReLU activation function. The number of feature maps are also doubled at each step. Image representations are progressively extracted in the contracting path, which also increases the dimension of the representations layer by layer. Densely connected convolutions~\cite{huang2016} was proposed to mitigate the problem of learning redundant features in successive convolutions for the U-net. In each step of the decoding path, they begin with an up-sampling function over the output of the previous layer. The feature maps are processed with bi-directional convolutional LSTMs~\cite{song2018} (BConvLSTM) unlike the original U-Net, where the corresponding feature maps in the contracting path are cropped and pasted onto the decoding path. 

After each up-sampling procedure, the outputs undergo batch normalization, which increases the stability of the network by standardizing the inputs to the network layer, by subtracting the batch mean and dividing the result by the batch standard deviation. Batch normalization helps to speed up training of the neural network. After batch normalization, the outputs are taken in by a BConvLSTM layer. The BConvLSTM layer uses two convolutional LSTMs~\cite{shi2015} (ConvLSTM) to process input data into forward and backward paths, from which a decision is made for the present input by handling the data dependencies in both directions. Note that in the original ConvLSTM, only the data dependencies for the forward path are processed. 

\section{Proposed Methods}

\subsection{Motivations}

This work was mainly inspired by BCDU-Net~\cite{azad2019} and LadderNet~\cite{laddernet}. The network adopts the general architecture of LadderNet, with components that of BCDU-Net. Like BCDU-Net, it adopts densely-connected convolutions and utilizes bi-directional convolutional LSTMs~\cite{song2018} (BConvLSTM), which has two convolutional LSTMs~\cite{shi2015} (ConvLSTM) to process input data in two directions (forward and backward paths), from which it makes a decision for the current input by handling data dependencies in both directions. 

Another inspiration for this work was from NAS-Unet~\cite{nasunet2019}. In neural architecture search for semantic segmentation for medical imaging, to the best of our knowledge, there has yet to be any published work with methods involving ensembles of U-Nets or its variants. A side-objective of this work is thus to illustrate the gains of such ensembles, using recent state-of-the-art methods in segmentation. In particular, this work shows that an ensemble of the same underlying network with varying sizes do yield better results than both with the same sizes. However, we were only able to experiment successfully with an ensemble of two networks and not more than that due to compute power limitations. 

\subsection{BUSU-Net}

The original proposed model - Big-U Small-U Net (BUSU-Net) consists of 108 layers made from chaining two BCDU-Nets, where the first is deeper than the second. In particular, the Big-U is deeper than the original BCDU-net~\cite{azad2019} and the Small-U is of the same size as the original. The visualization of the network can be found in Figure~\ref{busu-net} of the appendix (as it is really big). 

\subsection{LightBUSU-Net}

The LightBUSU-Net is a `lighter' version of BUSU-Net with 43 layers, where its Big-U and Small-U are less deep than in BUSU-Net and BCDU-Net. The purpose of this is to illustrate the robustness of a lightweight model, which can be deployed in situations with limited resources such as memory, space and compute capabilities. The visualization of the network can be found in Figure~\ref{lightbusu-net} of the appendix. 

\section{Experiments}

\subsection{Training of Neural Network}

All training was performed in the High Performance Computing\footnote{See \url{https://nusit.nus.edu.sg/hpc/} for more details.} (HPC) at the National University of Singapore (NUS). We utilized TensorFlow~\cite{tensorflow2016} version 1.12 in Python 3.6, 5 CPU cores, 100GB RAM, 1 unit of NVIDIA® Tesla® V100-32GB GPU. An example of the job script we submitted to the HPC cluster for training of the neural network is as follows:
\begin{lstlisting}[basicstyle=\scriptsize]
#!/bin/bash
#PBS -P volta_pilot
#PBS -j oe
#PBS -N BUSU_Net_experiment
#PBS -q volta_gpu
#PBS -l select=1:ncpus=5:mem=100gb:ngpus=1:mpiprocs=2
#PBS -l walltime=70:00:00
cd $PBS_O_WORKDIR;
np=$(cat ${PBS_NODEFILE} | wc -l);
image="/app1/common/singularity-img/3.0.0/tensorflow
_1.12_nvcr_19.01-py3.simg"
singularity exec "$image bash << EOF > stdout.$PBS_JOBID 
2> stderr.$PBS_JOBID"
#NCCL_DEBUG=INFO ; export NCCL_DEBUG
#mpirun -np $np -x NCCL_DEBUG python3 train_BUSU_Net.py
python3 train_BUSU_Net.py
EOF
\end{lstlisting}

\subsection{Evaluation Metrics}

We utilized several metrics to evaluate the performances of BUSU-Net and LightBUSU-Net, namely accuracy, sensitivity, specificity and F1-score. We first calculated the True Positive (TP), True Negative (TN), False Positive (FP) and False Negative (FN). The above metrics are thus calculated as follows: 
\begin{equation*}
    \text{accuracy} = \frac{\text{TP} + \text{TN}}{\text{TP} + \text{TN} + \text{FP} + \text{FN}}
\end{equation*}
\begin{equation*}
	\text{sensitivity} = \frac{\text{TP}}{\text{TP} + \text{FN}}
\end{equation*}
\begin{equation*}
    \text{specificity} = \frac{\text{TN}}{\text{TN} + \text{FP}}
\end{equation*}
\begin{equation*}
    \text{F1-score} = 2 \times \frac{\text{Precision} \times \text{Recall}}{\text{Precision} + \text{Recall}}
\end{equation*}
where 
\begin{equation*}
\text{Precision} = \frac{\text{TP}}{\text{TP} + \text{FP}},\quad \text{Recall} = \frac{\text{TP}}{\text{TP} + \text{FN}}
\end{equation*}

In particular, we calculated the receiver operating characteristics (ROC) curve and the area under curve (AUC) to further evaluate the performances of the neural networks. 

\begin{table*}[!htbp]
	\centering\caption{Performance comparison of proposed networks and recent state-of-the-art methods on DRIVE dataset}
	\resizebox{0.8\textwidth}{!}{%
		\begin{tabular}{cccccccc}
			\hline
			\textbf{Method}        & \textbf{Year} & \textbf{Dataset} & \textbf{Accuracy} & \textbf{Sensitivity} & \textbf{Specificity} & \textbf{AUC} & \textbf{F1-Score} \\ \hline
			COSFIRE filters~\cite{azzopardi2015}        & 2015              & DRIVE            & 0.9442            & 0.7655               & 0.9705              & 0.9614       & -                 \\
			Cross-Modality~\cite{li2016}         & 2015              & DRIVE            & 0.9527            & 0.7569               & 0.9816               & 0.9738       & -                 \\
			U-net~\cite{olaf2015}                  & 2015              & DRIVE            & 0.9531            & 0.7537               & \textbf{0.9820}                & 0.9755       & 0.8142            \\
			DeepModel~\cite{liskowski2016}              & 2016              & DRIVE            & 0.9495            & 0.7763               & 0.9768               & 0.9720        & -                 \\
			RU-Net~\cite{alom2018}                 & 2018              & DRIVE            & 0.9553            & 0.7726               & \textbf{0.9820}                & 0.9779       & 0.8149            \\
			R2U-Net~\cite{alom2018}                & 2018              & DRIVE            & 0.9556            & 0.7792               & 0.9813               & 0.9782       & 0.8171            \\
			LadderNet~\cite{laddernet}              & 2018          & DRIVE            & \textbf{0.9561}            & 0.7856               & 0.9810                & 0.9793       & 0.8202            \\
			BCDU-Net (d=1)~\cite{azad2019}         & 2019          & DRIVE            & 0.9559            & 0.8012               & 0.9784               & 0.9788       & 0.8222            \\
			BCDU-Net (d=3)~\cite{azad2019}        & 2019          & DRIVE            & 0.9560             & 0.8007               & 0.9786               & 0.9789       & 0.8224            \\ 
			Mi-UNet~\cite{sunet2019}      & 2019          & DRIVE            & 0.9559             & 0.8099               & 0.9772               & \textbf{0.9799}       & 0.8231            \\ \hline
			\textbf{LightBUSU-Net} & 2020          & DRIVE            & 0.9539            & \textbf{0.8281}               & 0.9723               & 0.9781       & 0.8207            \\
			\textbf{BUSU-Net}      & 2020          & DRIVE            & 0.9560             & 0.8113               & 0.9771               & \textbf{0.9799}       & \textbf{0.8243}            \\ \hline
		\end{tabular}%
	}
	
	\label{DRIVE-results}
\end{table*}

\section{Results}

We evaluated BUSU-Net and LightBUSU-Net on the DRIVE~\cite{drive2004} dataset. DRIVE is a dataset for blood vessel segmentation from retina images, and it consists of 40 color retina images, 20 of which are used for training and the remaining 20 images for testing. The original size of each image is $565 \times 584$ pixels. As it is clear that the number of samples in this dataset is not sufficient for training a deep neural network, we employ the same strategy as~\cite{alom2018,azad2019}. Firstly, the input images are randomly divided into patches numbering around $190,000$ from the 20 training images, of which $170,000$ are used for training, and the remainder $19,000$ patches are used for validation. 

Table~\ref{DRIVE-results} shows the quantitative results of the segmentation obtained from the recent state-of-the-art networks and that of the proposed networks on the DRIVE dataset. From the results, it can be observed that BUSU-Net outperforms the state-of-the-art networks on majority of the evaluation metrics, and has some very close to them. However, it is noteworthy to point out that LightBUSU-Net outperforms all state-of-the-art networks in sensitivity, including that of BUSU-Net. 

\begin{figure}[!htbp]
	\centering
	\includegraphics[scale=0.25]{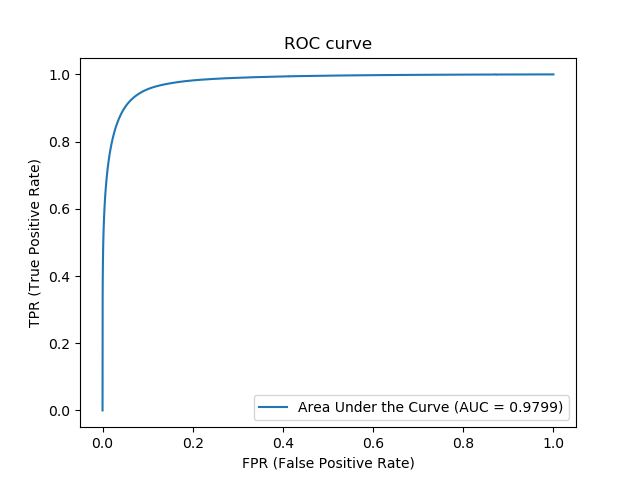}
	\includegraphics[scale=0.25]{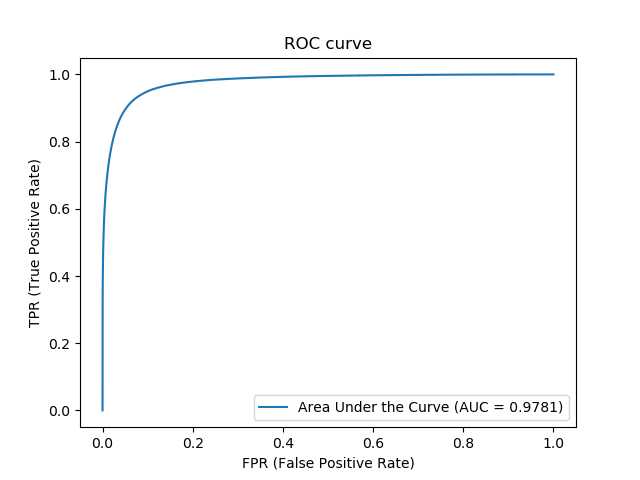}
	\caption{ROC curves of BUSU-Net (left) and LightBUSU-Net (right)}
	\label{ROC}
\end{figure}

\begin{figure}[!htbp]
	\centering
	\includegraphics[scale=0.25]{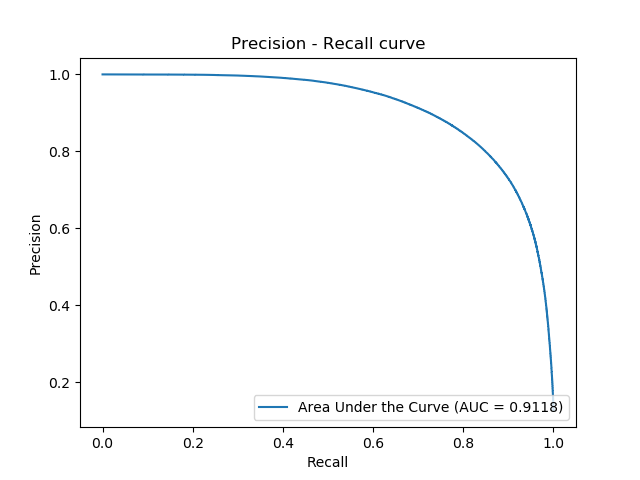}
	\includegraphics[scale=0.25]{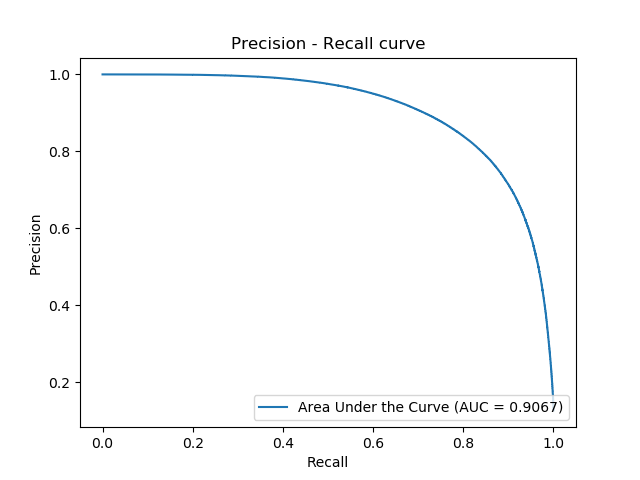}
	\caption{Precision recall curves of BUSU-Net (left) and LightBUSU-Net (right)}
	\label{Precision-recall}
\end{figure}

We illustrate the overall performances of BUSU-Net and LightBUSU-Net on the DRIVE dataset with their ROC curves, along with their precision recall curves in Figures~\ref{ROC} and~\ref{Precision-recall} respectively. The ROC curve is the plot of the true positive rate against the false positive rate. The area under the ROC curve (AUC), which is a measure of the network segmentation capability of the input data, can be found in Table~\ref{DRIVE-results}.

We also evaluated our results with LadderBCDU-Net, a network we designed as a benchmark which is made up of two original BCDU-Nets from~\cite{azad2019}. We did not present its results in Table~\ref{DRIVE-results} as it does not outperform the original BCDU-Net in any metric. Instead, we used it to illustrate the improvements in evaluation metrics when BCDU-Nets with different depths are superimposed to obtain BUSU-net. Interested readers may find LadderNet's implementation in the GitHub repository at~\url{https://github.com/juntang-zhuang/LadderNet}. Our implementation of BUSU-Net and LightBUSU-Net and their results which are presented in this paper can also be found at \url{https://github.com/weihao94/BUSU-Net}.


\section{Conclusion}

We have shown that our proposed BUSU-net has considerable gains as compared to recent state-of-the-art neural networks in semantic segmentation. In particular, we have made known that there are stark differences in terms of performance in a superimposed network of two BCDU-Nets of different depths as compared to a superimposed network of two BCDU-nets of equal sizes and also that of the original BCDU-Net. In other words, instead of having the same network of equal depths superimposed, by having the same network with unequal depths - one larger than the other, we can obtain overall better performance than that of a single network and that of two superimposed networks of equal depths. Furthermore, we have shown that a lightweight model of BUSU-Net is able to obtain much better results than recent state-of-the-art networks on some evaluation metrics. This, along with the results from BUSU-Net, will help in the work towards automated segmentation of medical images with an ensemble framework. 

\section{Future Work}

As long a stretch as it sounds, we will work towards automated segmentation of medical images, where ensembles of deep neural networks will be studied at a more rigorous level and generalized (possibly). There has been recent work on this like S-UNet~\cite{sunet2019}, a bridge-style U-Net architecture with a saliency mechanism and only 0.21 million parameters. It uses a minimal U-Net (Mi-UNet) architecture that significantly reduces the number of parameters to 0.07 million, compared to 31.03 million in the original U-Net. As future work, we hope to be able to deploy BUSU-Net (after parameter reduction methods) in an ensemble framework like S-UNet. In particular, the ensembles need not be of equal sized single U-Net, but rather of varying depths. Ensembles of BUSU-Nets with big-Us and small-Us of different depths can be chained together (equivalent to a Mi-UNet block in the S-UNet). For example, suppose that we denote a big-U as $U_b$ and a small-U as $U_s$. Then possible ensembles can be: $U_b U_s U_b U_s \dots U_b U_s$, $U_b U_s U_s U_b\dots U_s U_b$, $U_s U_b U_s U_b \dots U_s U_b$, etc. We were not able to experiment with such cases due to resource constraints in the HPC and also the fact that it is a shared resource utilized by many other researchers, where a large number of users in a queue can result in our job being sent to the back of the queue, due to a low priority score attributed to large amounts of compute resource being requested. If memory-efficient ensemble methods can be employed, it may become possible to work on an automated ensemble framework which combines both the ensembling framework above and that of NAS-unet. 

\section{Acknowledgments}

We will like to thank National University of Singapore's Information Technology for making GPUs readily available in the university and for their detailed yet user-friendly documentations. 

\nocite{*}
\bibliographystyle{plain}
\bibliography{references}

\appendix

The model visualizations of BUSU-Net and LightBUSU-Net can be found here, in the sections of the appendix tat follow. As the networks are quite deep (108 layers for BUSU-Net and 43 layers for LightBUSU-Net), their architectures span over the next few pages in the appendix. 

\section{BUSU-Net Visualization}

\begin{figure*}[!htbp]
	\centering
	\includegraphics[width=\textwidth,height=\textheight,keepaspectratio]{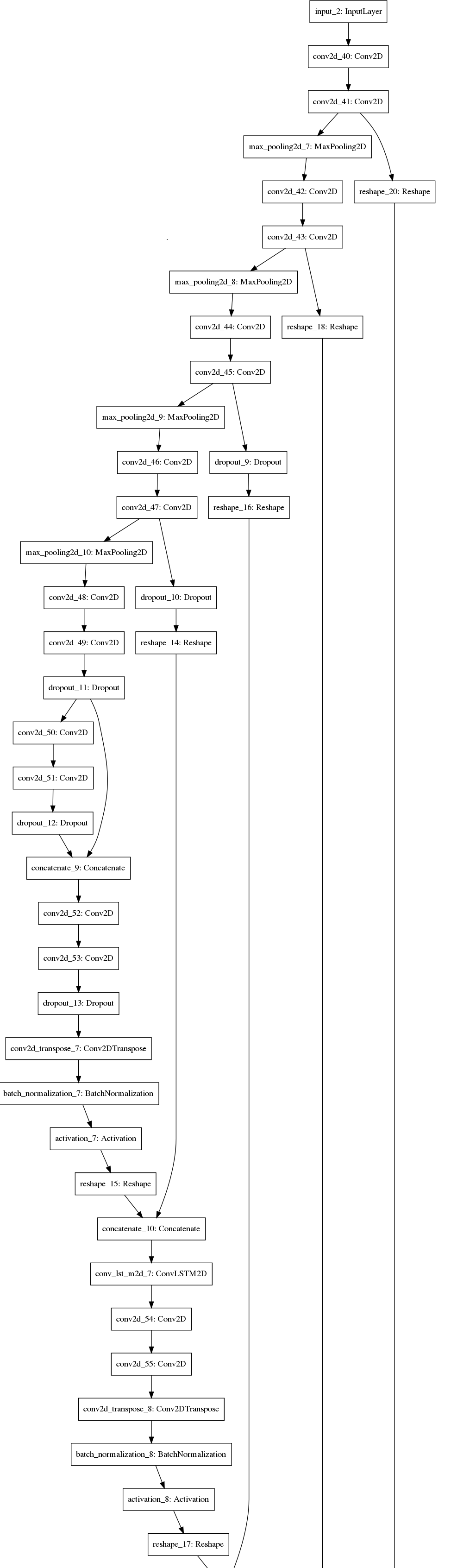}
\end{figure*}
\begin{figure*}[!htbp]
	\centering
	\includegraphics[width=\textwidth,height=\textheight,keepaspectratio]{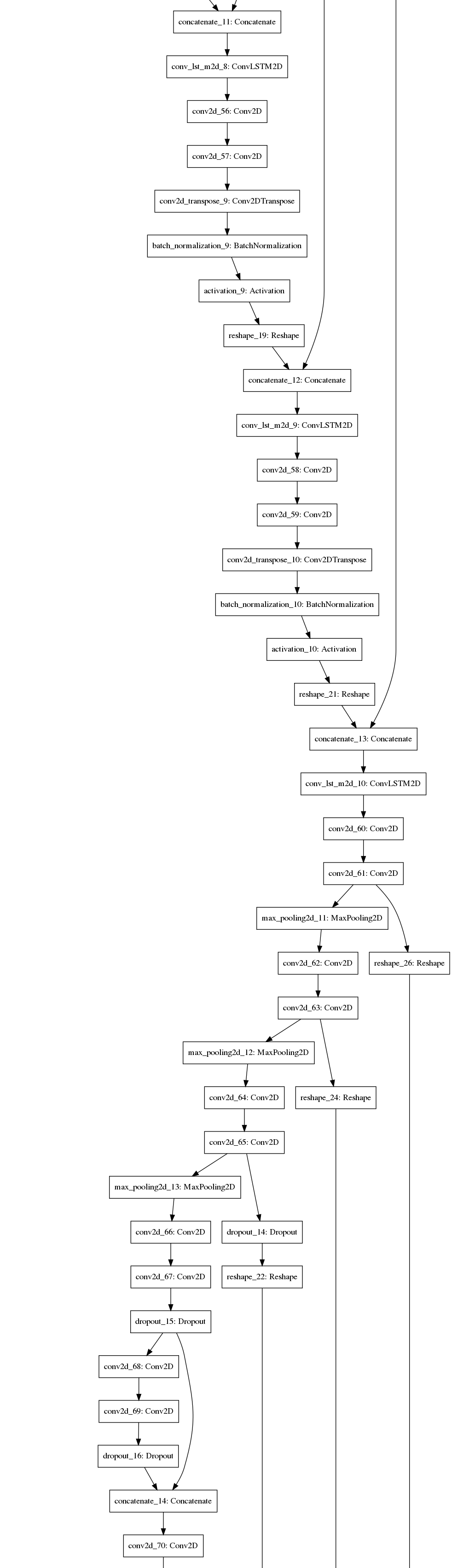}
\end{figure*}
\begin{figure*}[!htbp]
	\centering
	\includegraphics[width=0.8\textwidth,height=0.8\textheight,keepaspectratio]{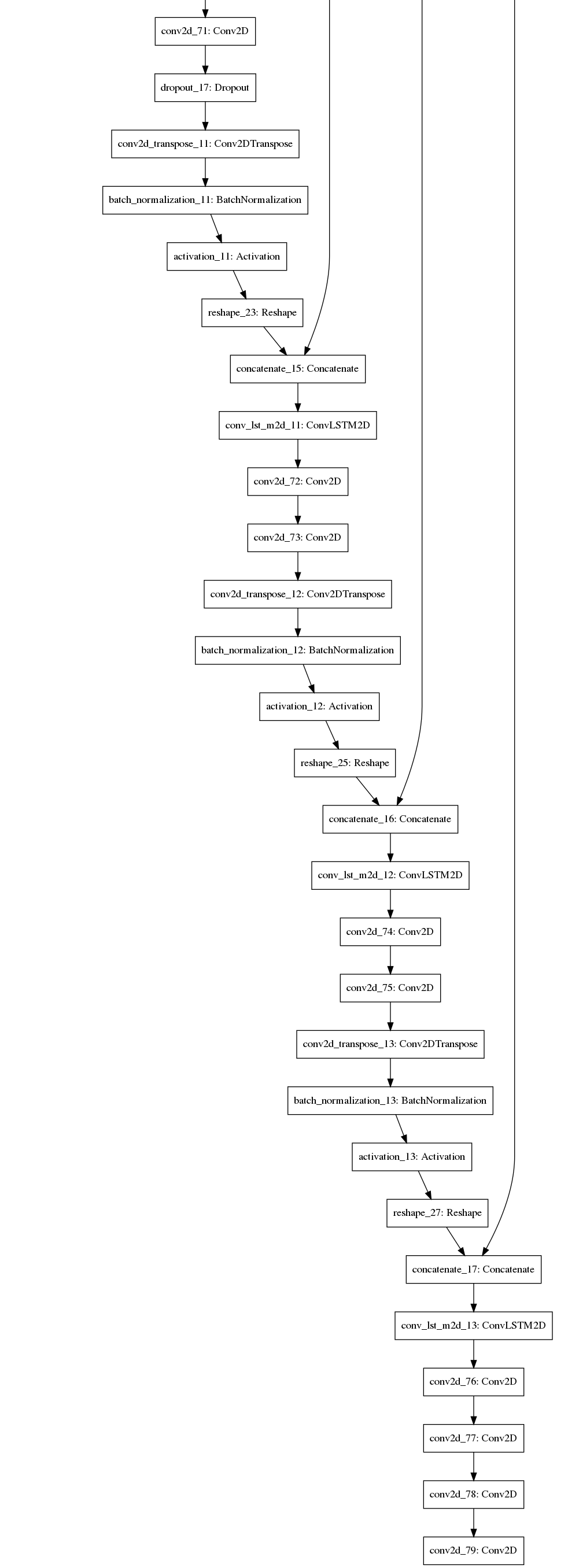}
	\caption{BUSU-Net}
	\label{busu-net}
\end{figure*}
\FloatBarrier

\section{LightBUSU-Net Visualization}

\begin{figure*}[!htbp]
	\centering
	\includegraphics[width=\textwidth,height=\textheight,keepaspectratio]{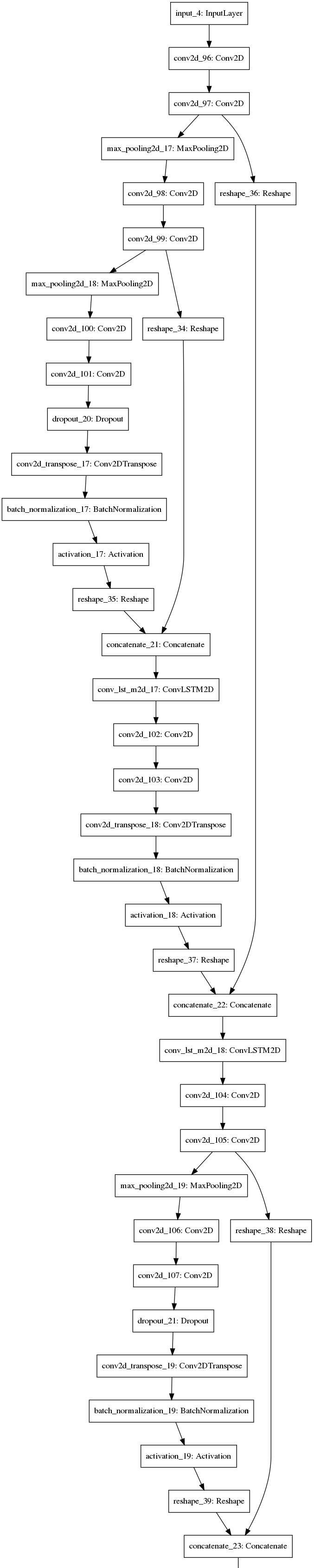}
\end{figure*}
\begin{figure*}[!htbp]
	\centering
	\includegraphics[width=0.3\textwidth,height=0.3\textheight,keepaspectratio]{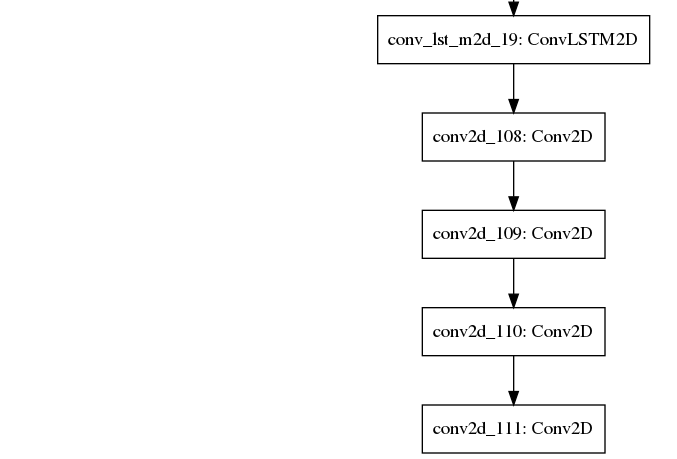}
	\caption{LightBUSU-Net}
	\label{lightbusu-net}
	\vspace{100in}
\end{figure*}
\FloatBarrier

\end{document}